# Laves Phase Crystal Analysis (LaCA): Atomistic Identification of Lattice Defects in C14 and C15 Topologically Close-Packed Phases

Zhuocheng Xie[a], Dimitri Chauraud[b], Erik Bitzek[b], Sandra Korte-Kerzel[a],

Julien Guénolé[a,c,d]

[a] Institute of Physical Metallurgy and Materials Physics, RWTH Aachen University, 52056 Aachen, Germany
[b] Department of Materials Science & Engineering, Institute I: General Materials Properties, Friedrich-Alexander-Universität Erlangen-Nürnberg, 91058 Erlangen, Germany
[c] Université de Lorraine, CNRS, Arts et Métiers ParisTech, LEM3, 57070 Metz, France
[d] Labex Damas, Université de Lorraine, 57070 Metz, France

## Abstract

The identification of defects in crystal structures is crucial for the analysis of atomistic simulations. Many methods to characterize defects that are based on the classification of local atomic arrangement are available for simple crystalline structures. However, there is currently no method to identify both, the crystal structures and internal defects of topologically close-packed (TCP) phases such as Laves phases. We propose a new method, Laves phase Crystal Analysis (LaCA), to characterize the atomic arrangement in Laves crystals by interweaving existing structural analysis algorithms. The new method can identify the polytypes C14 and C15 of Laves phases, typical crystallographic defects in these phases, and common deformation mechanisms such as synchroshear and non-basal dislocations. Defects in the C36 Laves phase are detectable through deviations from the periodic arrangement of the C14 and C15 structures that make up this phase. LaCA is robust and extendable to other TCP phases.

## 1. Introduction

Atomistic simulations are routinely used to study the deformation mechanisms of materials at the atomic scale [1–3]. Fundamental to the analysis of such simulations is the ability to identify defects in crystalline structures [4–7]. Many structure characterization methods, such as centrosymmetry parameter (CSP) [8] and common neighbor analysis (CNA) [9,10], are widely used for highly symmetric crystalline structures such as face-centered cubic (FCC), body-centered cubic (BCC) and hexagonal close packed (HCP) crystals. However, to the best of our knowledge, there is currently no method dedicated to the characterization of the local structural environments of topologically close-packed (TCP) phases, in particular of Laves phases.

Laves phases with ideal chemical composition $AB_2$ are TCP structures with three common polytypes: cubic $MgCu_2$ (C15), hexagonal $MgZn_2$ (C14) and $MgNi_2$ (C36) [11,12]. The larger A-type atoms are ordered as in cubic and hexagonal diamond structures and the smaller B-type atoms form tetrahedra around the A-type atoms. The ideal atomic size ratio for Laves phase formation is 1.225 but encompasses a range from 1.05 to 1.68 [13,14]. Laves phases have a layered structure which consists of quadruple atomic layers along the basal and {111} planes for the hexagonal and cubic Laves phases, respectively (see Figure 1a). The quadruple atomic layers consist of a single layer of B-type atoms forming a Kagomé net and a triple layer with an A-B-A structure [15,16]. The atomic distances within the triple layer are much smaller than the distance between the single and triple layers. The different compositions and stacking orders of the quadruple layers lead to the various polytypes of Laves phases.

Laves phases form in many alloys and have a large impact on their mechanical properties [12,17–19]. The understanding of the underlying deformation mechanisms is thus crucial for tailoring material properties. The synchroshear mechanism of the basal slip was predicted by Hazzledine *et al*. [15] for Laves phases, which was confirmed by their experimental observation of synchro-Shockley dislocation in the C14 $HfCr_2$ Laves phase [16] and by the *ab-initio* calculations performed by Vedmedenko *et al*. [20]. After synchroshearing a C14 Laves phase on the alternate triple layer, a strip of C15 structure is formed as a stacking fault. Additionally, Schulze and Paufler *et al*. [21–23] first reported non-basal slip in the C14 $MgZn_2$ Laves phase. Zehnder *et al*. [24] also observed the activation of non-basal slip and measured the corresponding critical resolved shear stresses in the C14 $Mg_2Ca$ Laves phase. Zhang *et al*. [25,26] revealed slip mechanisms for prismatic and pyramidal dislocations in a C14 Laves phase ($M_2Nb$, $M$=Cr, Ni, and Al).

In addition to those experimental studies and *ab-initio* calculations, the deformation of Laves phases was also investigated using classical atomistic simulations. This was in part enabled by the recently developed modified embedded atom method (MEAM) potential by Kim *et al*. [27] that accurately describes the mechanical properties of the C14 $Mg_2Ca$ Laves phase. Using this potential, synchroshear was identified as the most favorable mechanism for basal slip in the C14 $Mg_2Ca$ Laves phase [28]. This potential was also used to study the mechanisms of slip transmissions from an Mg matrix to the C14 $Mg_2Ca$ Laves phase at high temperatures using molecular dynamics (MD) simulations [29]. However, in these studies, the defects had to be identified using the von Mises shear strain invariant based on atomic-level strain tensors [30], as commonly used structural characterization methods such as CNA [9,10], CSP [8] and Voronoi index [31,32], are not able to identify Laves phase crystal structures or to detect their crystallographic defects. Hammerschmidt *et al*. [33] recently developed a method to characterize the crystal structures of TCP phases, including Laves phases, by computing the moments of the electronic density-of-states with analytic bond-order potentials on density-functional theory relaxed structures. Their approach is, however, not targeted towards identifying crystalline defects and is limited to systems with few hundreds of atoms.

With the aim of providing a better access to the deformation mechanisms of TCP phases via atomistic simulations, we chose Laves phases as archetype for TCP phases to develop a new structural analysis method named Laves phase Crystal Analysis (LaCA). By classifying the atoms according to the neighbor list obtained from the combination of existing structural characterization methods, the different Laves phase crystal structures and their typical crystallographic and chemical defects, e.g., anti-site atoms, stacking faults and dislocations, can be identified. In Section 2, we outline the workflow of the new analysis method. In Section 3, we assess the method with dedicated atomistic simulations of Laves phases and comparisons with existing analysis methods. The details of the implementation and the simulation methods are introduced in the Methodology section.

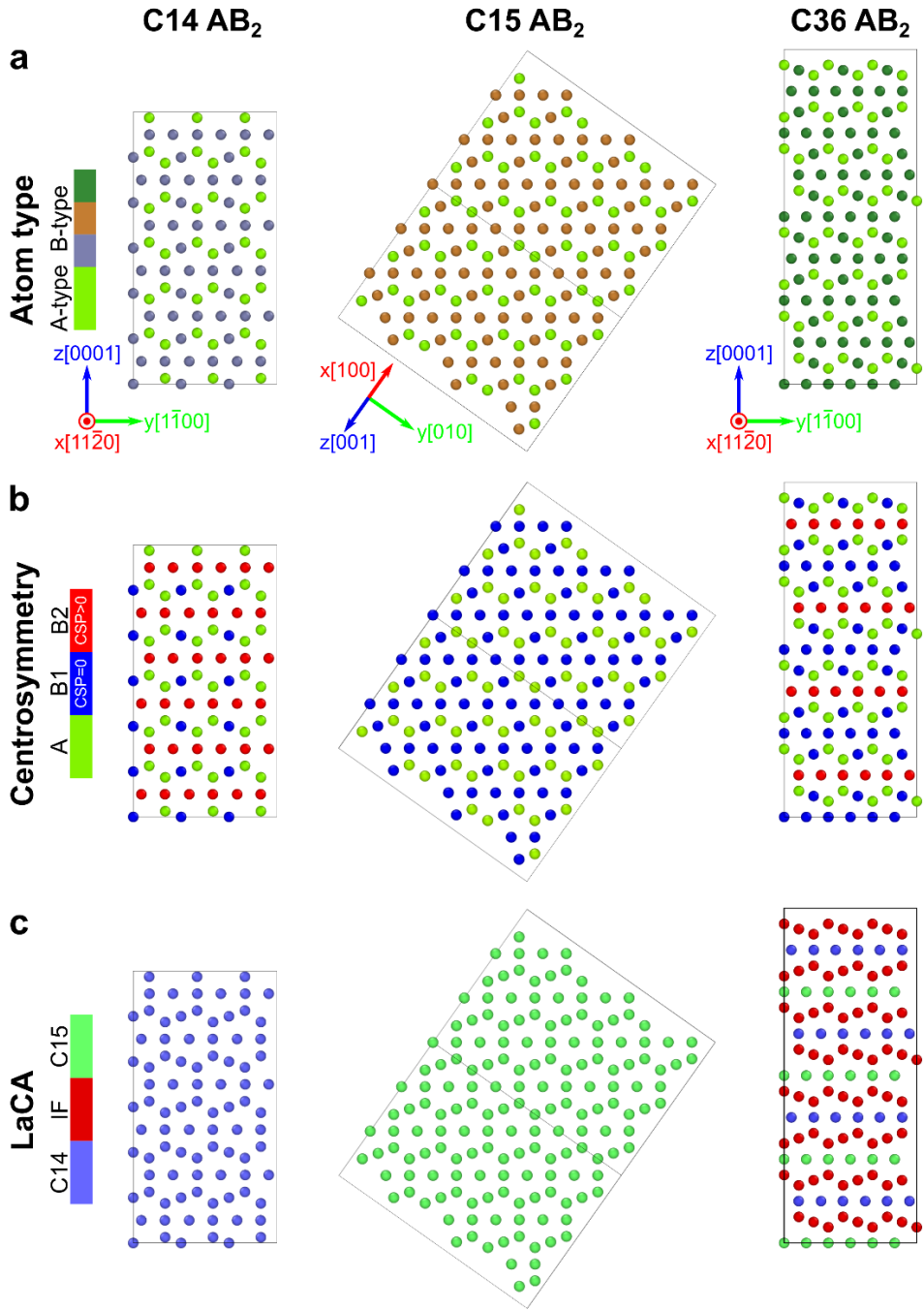

Figure 1: Laves phase crystal structures C14 (3×3×3 unit cells), C15 (3×3×3 unit cells) and C36 (3×3×2 unit cells). **a** Colored by particle type. A-type and B-type atoms are colored bright green and different subdued shades, respectively. **b** Colored by centrosymmetry parameter (CSP) which is calculated on the central atoms of Z12 Frank-Kasper (FK) clusters. B1-type and B2-type atoms are colored blue and red, respectively.

**c** Colored by Laves phase Crystal Analysis (LaCA) method. Atoms in C14, interface (IF) between C14 and C15, and C15 Laves crystal structures are colored blue, red, and green, respectively.

## 2. Laves phase crystal analysis

### 2.1. Identification of atomic clusters

To characterize the local structural environment of Laves phases, the CNA is used to identify the characteristic atomic clusters. The CNA (also known as the Honeycutt-Andersen bond type [9]) describes the number of shared neighboring atoms and the bonding relationships among the shared neighbors of a pair of atoms. The local environment of a pair of atoms is classified by three indices, $jkl$, where $j$ indicates the number of neighboring atoms the paired atoms have in common, $k$ indicates the number of bonds among the shared neighboring atoms, and $l$ indicates the number of bonds in the longest bond chain formed by the $k$ bonds among the shared neighbors. For instance, the only bonded pairs in a FCC crystal are 12× type 421, and a HCP crystal has equal numbers (6×) of type 421 and type 422. This has been widely used to identify FCC [34,35], HCP [36,37] and BCC [38,39] crystal structures. Laves crystal structures (C14, C15 and C36) consist of Z12 and Z16 Frank-Kasper (FK) clusters [40] as shown in Figure 2, in which the Z12 FK (icosahedral) cluster has 12× type 555 and the Z16 FK cluster has 12× type 555 and 4× type 666. A- and B-type atoms are the central atoms of Z16 and Z12 FK clusters in $AB_2$ Laves phases, respectively. We modified the implementation of the CNA in the Open Visualization Tool (OVITO) [41] to be able to identify Z12 and Z16 FK clusters in Laves phases. Details of the implementation of the modified CNA can be found in the Methodology section.

Alternatively, the Voronoi index can be used to determine the local atomic environment by identifying the Voronoi polyhedron enclosing an atom. A Voronoi polyhedron can be expressed using a vector of Voronoi indices ($n_1$, $n_2$, $n_3$, $n_4$, $n_5$, $n_6$, …), where $n_i$ is the number of polyhedron faces with i edges. The Voronoi cell of a central atom of the Z12 FK cluster has 12 faces with five edges each, its index vector is (0, 0, 0, 0, 12, 0). For a central atom of the Z16 FK cluster, the Voronoi cell has 12 faces with five edges each and 4 faces with six edges each, and the corresponding Voronoi index vector is (0, 0, 0, 0, 12, 4).

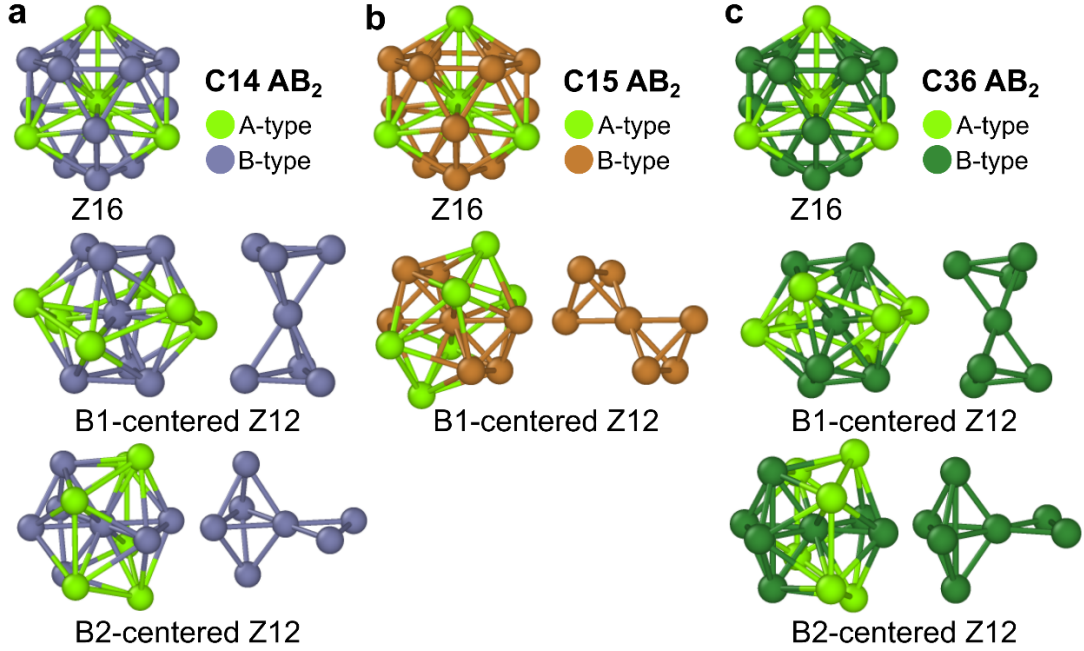

Figure 2: Characteristic atomic clusters in Laves phase crystal structures **a** C14, **b** C15 and **c** C36. B-type atoms in the Z12 icosahedral clusters are shown on the right side of the individual Z12 icosahedral clusters.

### 2.2. Centrosymmetry

The spatial and chemical distributions of neighboring atoms around the central atom of the Z16 FK cluster are the same in these three Laves phases. However, the local structural environment around the central atom of the Z12 FK cluster in C15 is different than in C14 and C36 Laves phase crystal structures. As shown in Figure 2, C14 and C36 Laves phases have two types of Z12 FK clusters with different chemical distributions of neighboring atoms around the central atoms. These two Z12 FK clusters are classified as B1- and B2-centered clusters (with central atoms B1 and B2). The C15 Laves phase only has B1-centered Z12 FK clusters. To differentiate B1- and B2-centered clusters mathematically, the centrosymmetry parameter (CSP) is used. By providing the number of nearest neighbors ($N$), the CSP is calculated by:

$$CSP = \sum_{i=1}^{N/2} |r_i + r_{i+N/2}|^2 \qquad (1)$$

where $r_i$ and $r_{i+N/2}$ are vectors corresponding to a pair of opposite nearest neighbors [8]. The ideal centrosymmetric crystal structure has a CSP value of zero since the contributions of all nearest neighboring pairs are canceled out. The CSP is calculated on B-type atoms with $N=6$ nearest neighbors (only B-type atoms are considered). B1-type atoms have a CSP value of zero due to the centrosymmetric structure and, in contrast, B2-type atoms have a CSP value larger than zero. As shown in Figure 1b, C14, C15 and C36 Laves crystal structures can be differentiated by color coding according to the CSP values of B1- and B2-type atoms.

### 2.3. Neighbor list

To characterize the Laves phase crystal structures quantitatively, neighbor lists of atoms in the characteristic structures are built by considering the type of atomic cluster and the CSP value (see

Table 1). The neighbor list is constructed by considering the number of nearest neighbors as defined by the modified CNA classification, e.g., for central atoms of the Z12 and Z16 FK clusters, 12 and 16 nearest neighbors are visited, respectively. The classifications include atoms in C14 (A-, B1- and B2-types), C15 (A- and B1-types), interface (IF) between C14 and C15 (A1-, A2- and B1-types), Other Laves (OL) and Other. For example, if a central atom (A-type) of one Z16 FK cluster has a neighbor list of 4 × A-type, 3 × B1-type, and 9 × B2-type neighboring atoms, this A-type atom is classified as C14 Laves phase. Figure 1c shows C14, C15 and C36 Laves phases characterized by neighbor listing. C14 and C15 Laves phases can be directly recognized, whereas the C36 Laves phase consists of a periodic succession of C14 and C15 structures.

In the classification of the neighbor list, small perturbations due to thermal fluctuation and elastic deformation are allowed. The neighbor list is treated as a vector where the numbers of A-, B1- and B2-type neighbors are the first, second and third components, respectively. The Euclidean distance between the referenced vector and the vector of interest is calculated:

$$d_{\text{Eucl}}(\boldsymbol{I}, \boldsymbol{R}) = \sqrt{\sum_{i=1}^{N}(\boldsymbol{R}_i - \boldsymbol{I}_i)^2} \qquad (2)$$

where $\boldsymbol{R}$ is the referenced vector and $\boldsymbol{I}$ is the vector of interest. If the Euclidean norm is below the threshold of similarity (see Table 1), the atom is classified as the referenced Laves crystal structure, otherwise the atom is classified as OL.

Table 1: Neighbor lists of atoms in Laves phase crystal structures: C14, C15 and interface (IF) between C14 and C15. Other Laves (OL) indicates an atom identified as the central atom of Z12 or Z16 FK clusters that does not match the referenced neighbor lists. Other indicates an atom not identified as the central atom of Z12 or Z16 FK clusters. The Euclidean distance between the referenced vector and the vector of interest is calculated. If the Euclidean distance is below the threshold of similarity, the atoms belong to the characteristic Laves crystal structures.

| Type | Atomic cluster | CSP of B-type atom | Referenced neighbor list | Referenced vector | Threshold of similarity |
|---|---|---|---|---|---|
| **C14, A** | Z16 | - | 4×A, 3×B1, 9×B2 | (4, 3, 9) | $d_{\text{Eucl}} \leq \sqrt{2}$ |
| **C14, B1** | Z12 | ≈0 | 6×A, 0×B1, 6×B2 | (6, 0, 6) | $d_{\text{Eucl}} \leq \sqrt{2}$ |
| **C14, B2** | Z12 | ≫0 | 6×A, 2×B1, 4×B2 | (6, 2, 4) | $d_{\text{Eucl}} = 0$ |
| **C15, A** | Z16 | - | 4×A, 12×B1, 0×B2 | (4, 12, 0) | $d_{\text{Eucl}} \leq \sqrt{2}$ |
| **C15, B1** | Z12 | ≈0 | 6×A, 6×B1, 0×B2 | (6, 6, 0) | $d_{\text{Eucl}} \leq \sqrt{2}$ |
| **IF, A1** | Z16 | - | 4×A, 6×B1, 6×B2 | (4, 6, 6) | $d_{\text{Eucl}} \leq \sqrt{2}$ |
| **IF, A2** | Z16 | - | 4×A, 9×B1, 3×B2 | (4, 9, 3) | $d_{\text{Eucl}} \leq \sqrt{2}$ |
| **IF, B1** | Z12 | ≈0 | 6×A, 3×B1, 3×B3 | (6, 3, 3) | $d_{\text{Eucl}} = 0$ |
| **OL** | Z12 or Z16 | - | Else | - | - |
| **Other** | Else | - | - | - | - |

## 2.4. Dislocation analysis

The Dislocation Extraction Algorithm (DXA) [42] is applied to atomistic configurations to extract dislocation lines and determine their Burgers vectors. In brief, the algorithm tries to build Burgers circuits around clusters of atoms that have been identified as crystallographic defects by the CNA method, assuming they could form dislocation cores. The implementation of the DXA in the OVITO can extract dislocations in FCC, HCP, BCC, diamond cubic, and diamond hexagonal crystal structures. For Laves phases, the central atoms of Z16 FK clusters (A-type atoms) are ordered as in diamond structures (diamond hexagonal for C14 Laves phase and diamond cubic for C15 Laves phase). Moreover, the crystallographic information of Laves phases can be solely defined based on the A-type atoms. Therefore, the dislocations and their Burgers vectors in Laves crystal structures can be extracted by performing the DXA on the central atoms of Z16 FK clusters which are classified using the modified CNA presented in this work.

## 2.5. Transferability of LaCA

The C36 hexagonal unit cell is made of a periodic stacking along the c axis of one C14 unit cell and one C15 unit cell. The identification of the C36 phase thus requires a larger cutoff able to include this periodic stacking when constructing the neighbor list. Any defects formed in the C36 phase will either break this stacking or directly create a defect in one of the two sub-phases. It remains however unclear how to precisely identify defects in the C36 phase, in particular when considering planar defects such as stacking faults or twins. For example, by considering the glide of a synchroshear partial dislocation in the C36 phase, that transforms a C14 layer into a C15 layer: is the resulting structure a C36 phase with a stacking fault, or a dual-phase composed of a C36 and C15 phase? A similar question can be raised with the glide of a synchroshear partial dislocation in the C14 phase: is the resulting defected C14 phase embedding a layer of C15 phase or a layer of C36 phase formed by the C15 layer and an adjacent C14? To our knowledge, no generally agreed-upon definition currently exists. A solution could be to adapt the results of LaCA based on the user input, whether the C36 phase has to be found in the sample or not. Such an approach would be similar to experimental analysis, where growth and deformation defects cannot always be distinguished based on a post-mortem analysis only, and as such requires prior knowledge on the sample. This is however beyond the development of the LaCA analysis method and thus out of the scope of the current work.

Besides the C36 Laves phase, LaCA can be extended to identify other TCP phases. Indeed, the crystallographic structures of any TCP phases can be classified into coordination polyhedra (denoted by FK clusters) by means of CNA or Voronoi index analysis. Then, the spatial and chemical distribution of the polyhedral units can be identified using the CSP and the neighbor listing. For each new TCP phase to be identified, the relationships between the atom structure types and the FK clusters have to be found, as in Table 1 in the case of Laves phases. Additionally, these parameters need to be adapted on a case-by-case basis:

- the ratio between the CNA adaptive cutoff and the first neighboring distance,
- the number of nearest neighbors when calculating the CSP value,
- the threshold of the CSP value to evaluate the centrosymmetry,
- the referenced neighbor list,

- the threshold of similarity.

## 3. Assessments

### 3.1. Identification of typical crystallographic defects

Typical crystallographic defects in Laves phases are summarized in Figure 3, as identified using the percentage deviation of potential energy to bulk ($\Delta E_\mathrm{p}$) and our method LaCA. Details of the simulation method employed to generate these atomistic defects can be found in the Methodology section. On the left side of each pair of images in Figure 3, coloring is based on $\Delta E_\mathrm{p}$, while the right side of each pair is colored using LaCA. Point defects such as anti-site atoms and vacancies are widely observed experimentally in Laves phases, especially in off-stoichiometric compositions and at elevated temperatures [12]. As shown in Figure 3a, the values of $\Delta E_\mathrm{p}$ around the vacancies are below 4.1% (~0.1 eV) in C14 $Mg_2Ca$. By applying LaCA, the atoms in the first two nearest neighbors of vacancies are identified as Other. The anti-site defects in C14 $Mg_2Ca$ are identified as shown in Figure 3b. For perfect Laves crystals, A-type atoms are the center atoms of Z16 FK clusters, and B-type atoms are the center atoms of Z12 FK clusters. As LaCA combines the results of the modified CNA and the particle type information from the atomistic configuration, chemical defects such as anti-site atoms and anti-phase boundaries can be identified.

In Laves phases, plastic deformation occurs via dislocation-mediated processes. Dislocation motion on the basal plane in the hexagonal and the {111} plane in the cubic Laves phases occurs by synchroshear, characterized by two synchro-Shockley dislocations moving on adjacent planes of the triple layer [16,20,28]. Figure 3c shows the core structures of two typical synchro-Shockley dislocations with mixed characters (60° and 30°) in C14 $Mg_2Ca$. Atoms at the dislocation cores show higher values of $\Delta E_\mathrm{p}$ up to 17.6% (~0.5 eV) than the atoms in stacking faults, and are identified as Other using LaCA. The stacking faults (C15 structure), which are bounded by the synchro-Shockley partial dislocations, are identified using LaCA, as well as the interface between the C15 stacking fault and the C14 matrix. Due to the low stacking fault energy of C14 $Mg_2Ca$ (14 mJ/m$^2$), the $\Delta E_\mathrm{p}$ of the atoms in stacking faults is less than 0.3% (~0.006 eV), see Figure 3d. The stacking fault and twin boundary in C15 $Mg_2Ca$ are also identified using LaCA. Moreover, atoms at the free surfaces are classified as Other with LaCA (Figure 3e).

Similarly, point defects, dislocation cores and free surfaces in C15 and C36 Laves phases can be identified by LaCA, since the local atomic packings near these defects are not classified as Z12 or Z16 FK clusters using the modified CNA.

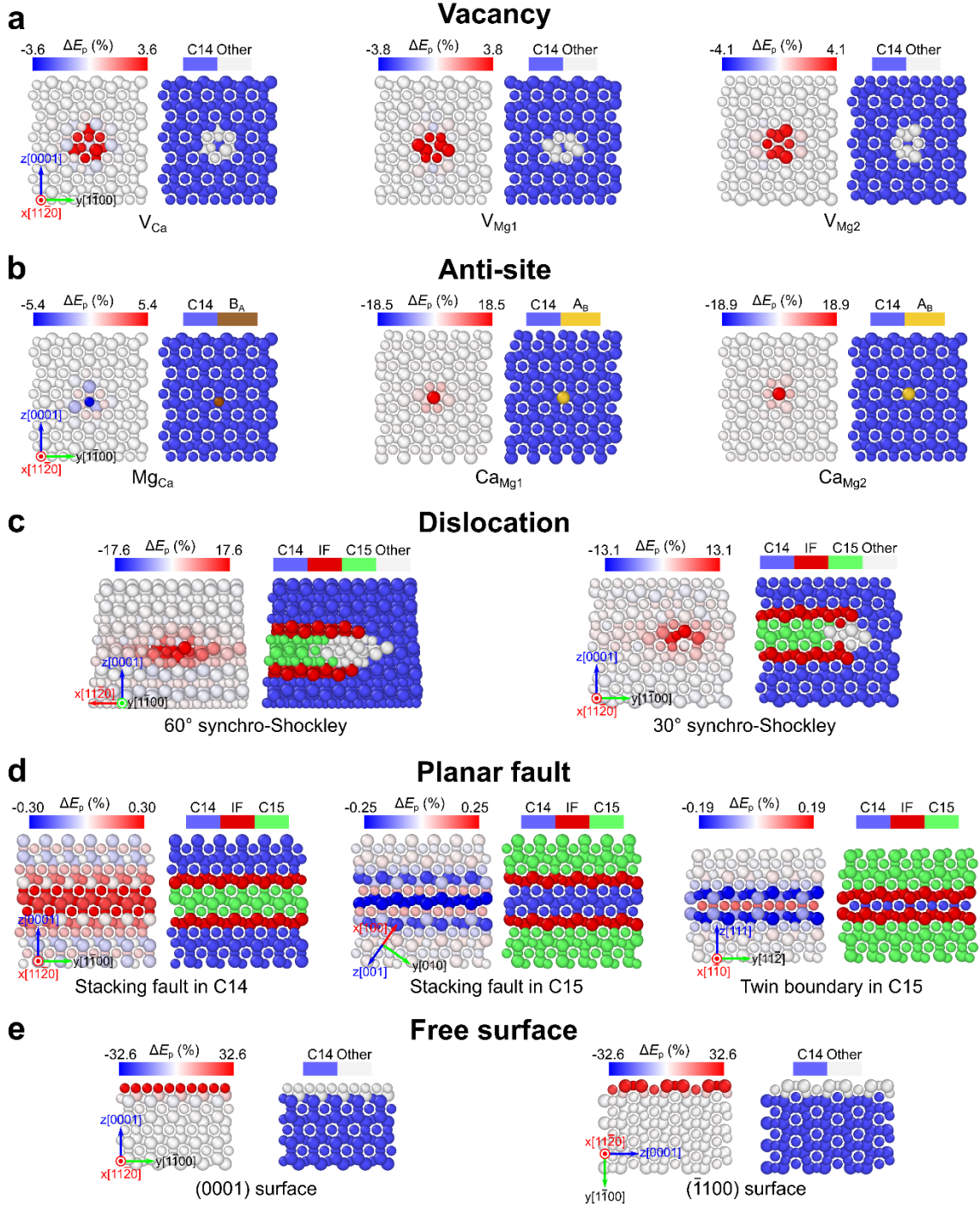

Figure 3: Overview of typical crystallographic defects in Laves phases. Left: colored by percentage deviation of potential energy to bulk ($\Delta E_p$). Right: colored by LaCA. Large and small atoms are A- and B-type atoms, respectively. The atomic size ratio is 1.231 for Mg-Ca systems. **a** Vacancy defects (denoted, e.g., as $V_A$ for vacancy on an A site) in C14 $Mg_2Ca$. **b** Anti-site defects (denoted, e.g., as $A_B$ for an A atom on a B site) in C14 $Mg_2Ca$. **c** Synchro-Shockley dislocations in C14 $Mg_2Ca$. **d** Stacking faults and twin boundary in C14 and C15 $Mg_2Ca$. **e** Free surfaces in C14 $Mg_2Ca$.

## 3.2. Identification of basal slip in the hexagonal C14 structure: the synchroshear mechanism

The synchroshear mechanism is the most favorable mechanism for basal slip in the C14 Laves phase. It has been confirmed by our previous study using nudged-elastic band (NEB) calculations [28] and experimentally in the Laves phase layer of the $\mu$-phase crystal structure [43,44]. It is likely to occur in the same way on the basal plane of the C36 phase and the {111} plane of the C15 phase. The minimum energy path (MEP) of a full slip with a Burgers vector of $1/3<11\bar{2}0>$ is associated with two energy maxima corresponding to two separated synchro-Shockley dislocations with Burgers vectors of $1/3<10\bar{1}0>$. Details of the simulations can be found elsewhere [28]. Figure 4 shows the propagation of the leading and the trailing synchro-Shockley dislocations, and the formation and annihilation of a stacking fault during the slip in a C14 $Mg_2Ca$ pillar are visualized using atomic shear strain, modified CNA, LaCA, and the combination of LaCA and DXA. The atomic shear strain characterizes the deformation gradient of the slip event without further information on the dislocation and stacking fault (Figure 4a-c(ii)). By applying the modified CNA, atoms belonging to dislocation cores and free surfaces are separated from the central atoms in Z12 and Z16 FK clusters (Figure 4a-c(iii)). The stacking fault with C15 Laves crystal structure, produced by the passage of the leading synchro-Shockley dislocation, is identified using LaCA, see Figure 4a-c(iv). By combining LaCA and DXA, the synchroshear mechanism is revealed with a dislocation line and a stacking fault (Figure 4a-c(v)). In addition, the Burgers vectors of the leading and trailing synchro-Shockley dislocations are correctly characterized as $1/3<10\bar{1}0>$.

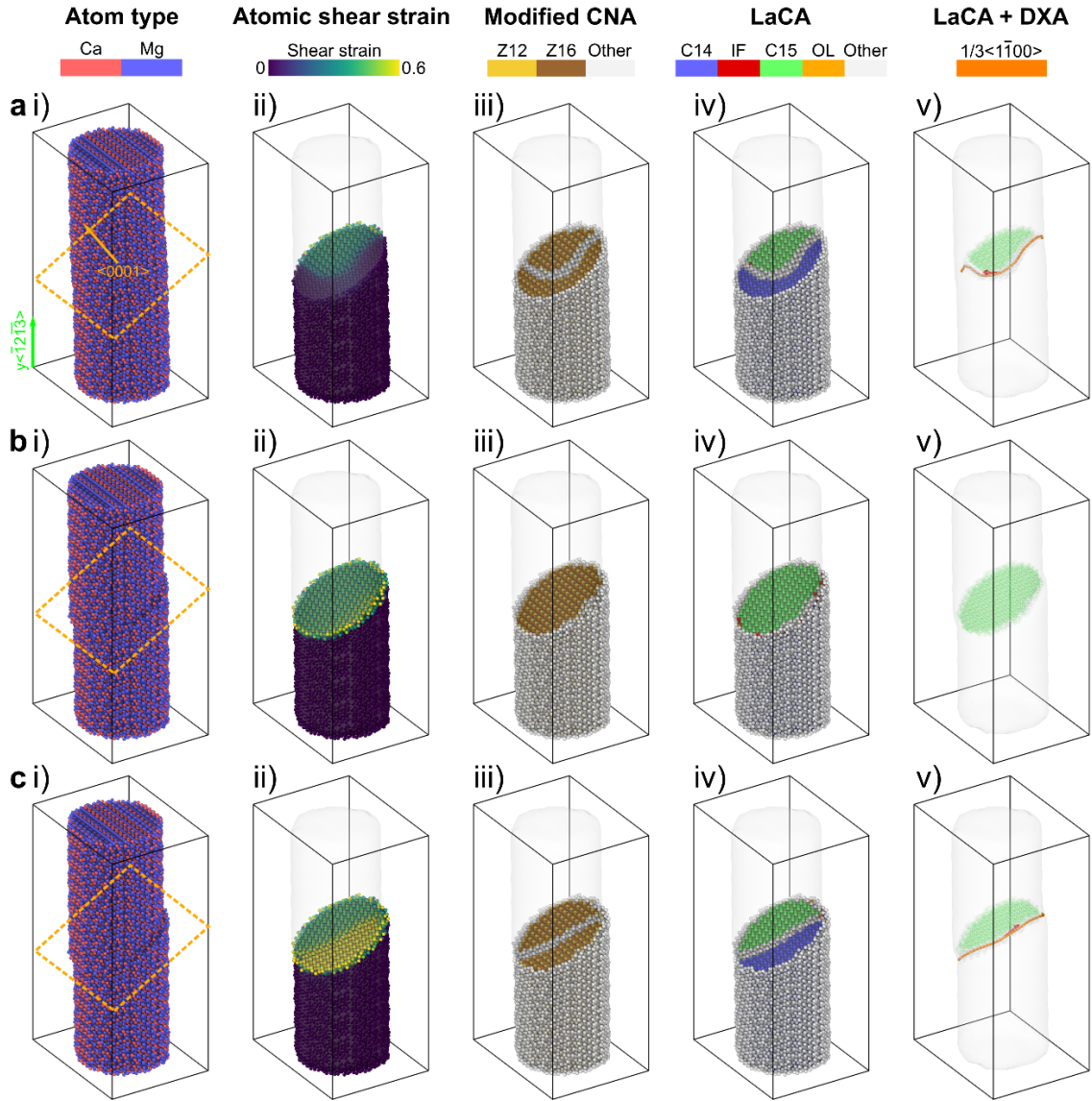

Figure 4: Synchroshear mechanism for basal slip in C14 Mg$_2$Ca Laves phase as computed by nudged-elastic band (NEB) method. **a** Glide of the leading synchro-Shockley dislocation. **b** Formation of a stacking fault (C15 Laves crystal structure) in the C14 Laves phase after the glide of the leading synchro-Shockley dislocation. **c** Glide of the trailing synchro-Shockley dislocation. i) Colored by particle type, the dashed box indicates the slice along the basal plane in ii)-v). ii) Colored by shear strain. iii) Colored by the modified CNA. iv) Colored by LaCA. v) DXA was performed on the Laves phase crystal structures as identified by the modified CNA, by considering only the central atoms of the Z16 atomic clusters. Only atoms in stacking faults (C15 Laves crystal structure and Other) are shown in half-transparent. Burgers vector is colored red. The surface mesh of the pillar is shown in half-transparent.

### 3.3. Identification of non-basal slip in the C14 structure: slip transmission at Laves phase/metal interfaces

Slip transmission at Laves phase interfaces is important to understand the strengthening effect of Laves precipitates in a metallic matrix. In our previous study, we simulated the co-deformation of Mg$_2$Ca C14 Laves phase with an $a$-Mg matrix by means of atomistic simulations in order to explore the transmission of dislocations at the interface [29]. The uniaxial compressive deformation was applied parallel to the matrix – Laves phase interface (along the y-axis in Figure 5) at low temperature within the microcanonical (NVE) ensemble (initial temperature at 0 K). More details on the simulations can be found elsewhere [29]. These atomistic configurations are of particular interest to assess the LaCA method, as they contain two phases, one interface and the nucleation of a non-basal dislocation in the Laves phase on a prismatic plane. Here, this prismatic slip in the Laves phase is analyzed using atomic shear strain, modified CNA, LaCA, and the combination of LaCA and DXA (Figure 5). The deformation gradient of the slip transfer is measured using atomic shear strain (Figure 5b). Mg matrix, Laves phase interface, and C14 $Mg_2Ca$ Laves phase are classified using the modified CNA. The stacking faults and twins in the Mg matrix and the potential dislocation cores in the C14 Laves phase are identified (Figure 5c). The crystal structure remains C14 after the dislocation glide according to LaCA, see Figure 5d, which indicates that the dislocation is a perfect dislocation. DXA on the configurations analyzed by the modified CNA with only the central atoms of Z16 FK clusters reveals the dislocation to possess a Burgers vector of $1/3<1\bar{2}10>$, corresponding to a perfect <$\alpha$>-dislocation on the prismatic slip plane (Figure 5e).

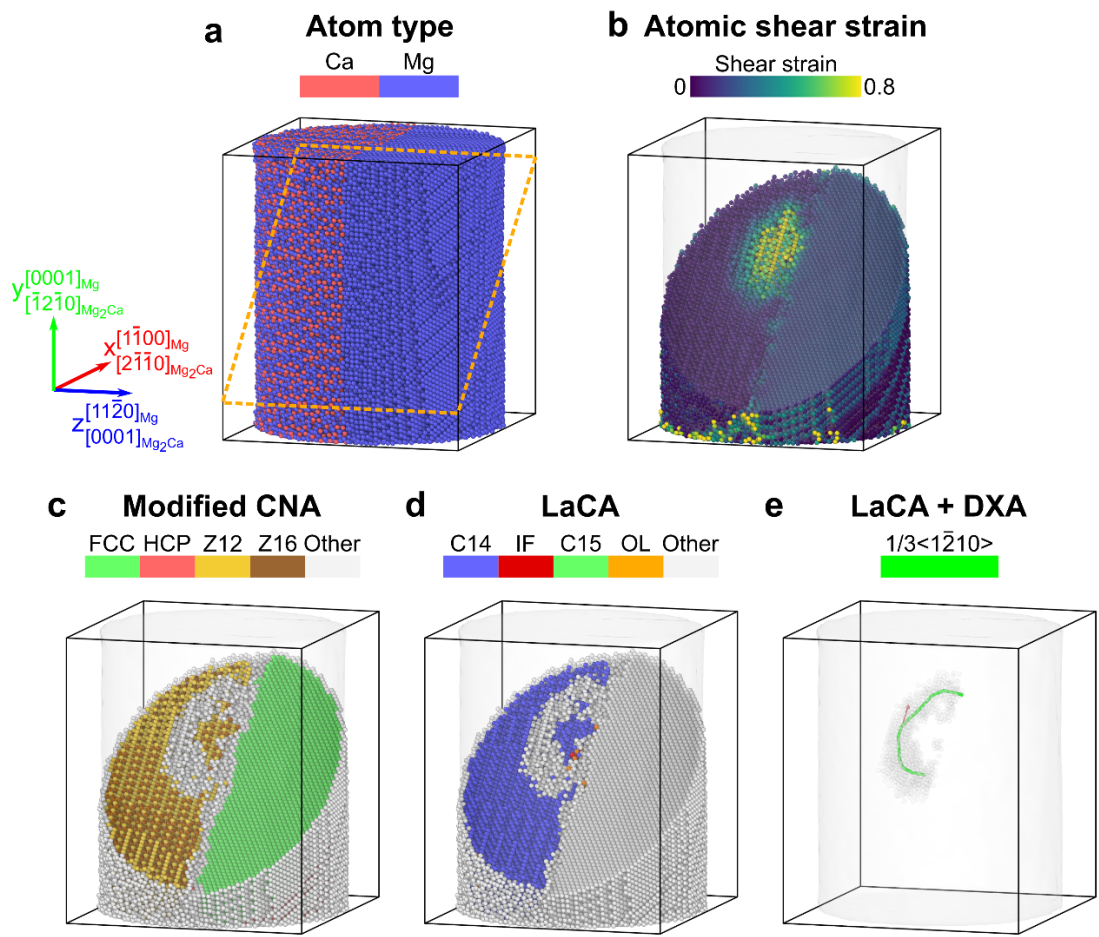

Figure 5: Slip transmission from the basal plane of the Mg matrix to the prismatic plane of the C14 $Mg_2Ca$ Laves phase during a compression test at 50 K using MD simulations. **a** Colored by particle type, the dashed box indicates the slice along the parallel basal (Mg matrix) and prismatic (C14 $Mg_2Ca$) planes in **b-e**. **b**

Colored by shear strain. **c** Colored by modified CNA. **d** Colored by LaCA. **e** DXA was performed on the post-processed Laves crystal structures using modified CNA, where only central atoms of Z16 atomic clusters were used. Only atoms in dislocation core (atoms identified as Other by LaCA) are shown in half-transparent. Burgers vector is colored red. The surface mesh of the sample is shown in half-transparent.

### 3.4. Identification of unknown defects: nanoindentation

Nanoindentation tests by MD simulations were performed on a $Mg_2Ca$ C14 Laves phase to explore activate plastic events with strain concentration, see Figure 6a. Details of the simulation method can be found in the Methodology section. The atomistic configuration of the $Mg_2Ca$ C14 Laves phase after nanoindentation is analyzed using atomic shear strain, relative potential energy to bulk $\Delta E_\mathrm{p}$, LaCA, and the combination of LaCA and DXA, see Figure 6b-d. As opposed to the post-processed results from shear strain and $\Delta E_\mathrm{p}$ filtering methods, LaCA successfully captures the overall plastic events under the localized strain. All high-energy atoms were correctly identified by LaCA as being part of a defect configuration as shown in Figure S1. Moreover, dislocation cores, stacking faults, vacancies and anti-site defects in the deformed sample are recognized using LaCA. Dislocations, such as <$a$> dislocations on basal and prismatic slip planes and synchro-Shockley dislocations, are characterized using DXA on the configuration identified with the modified CNA.

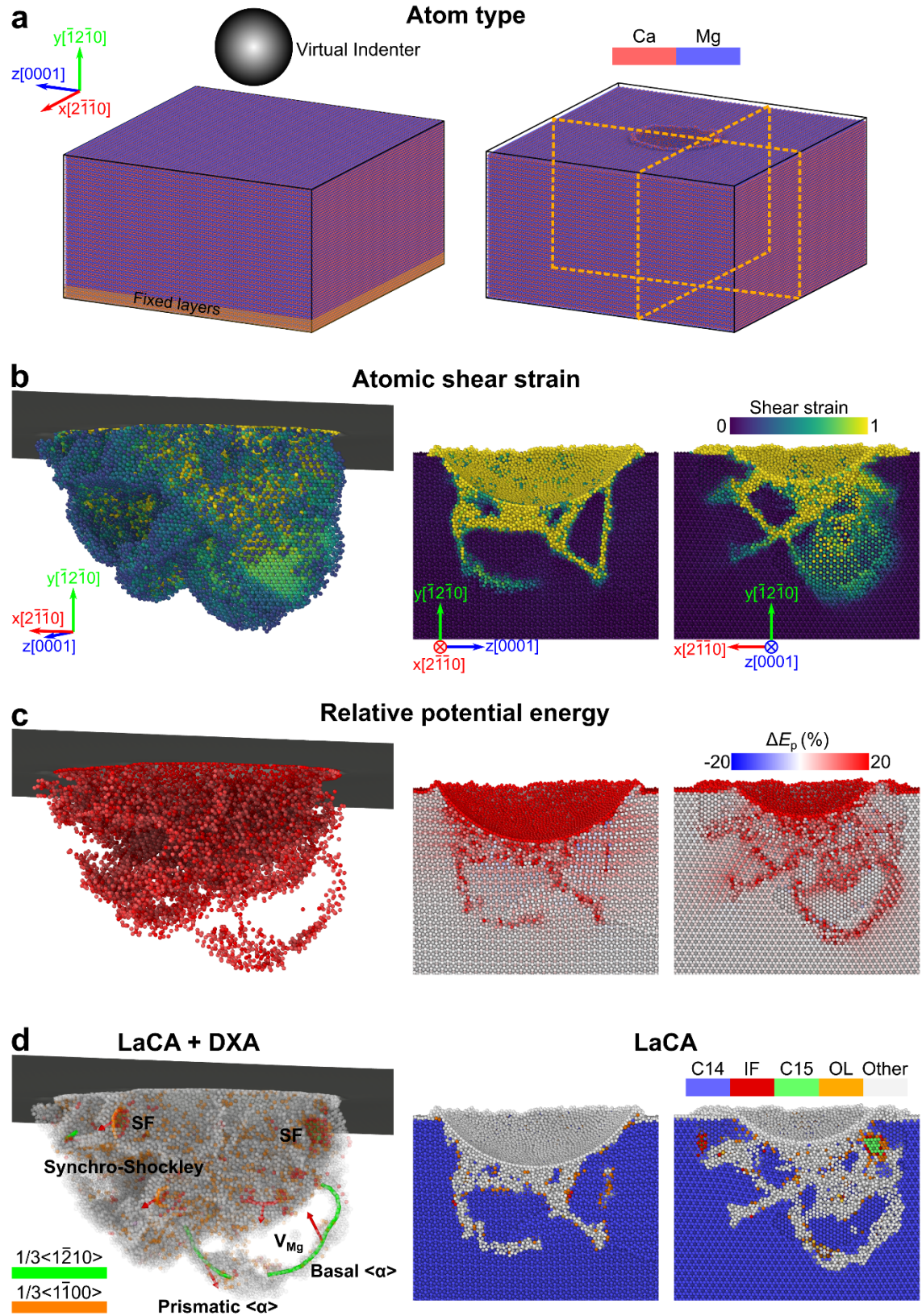

Figure 6: Nanoindentation test on the C14 Mg$_2$Ca Laves phase. **a** Atomistic sample before and after nanoindentation (indent depth = 5 nm) colored by particle type. The dashed boxes indicate the slices

perpendicular to the x and z axes in the right insets of **b-d**. **b-d** Plastic deformation zone beneath the indent. Colored by **b** shear strain (left: only atoms with shear strain > 0.2 are shown), **c** $\Delta E_p$ (left: only atoms with $|\Delta E_p|$ > 12% are shown), and **d** LaCA (left: atoms in C14 structure are not shown) and DXA (Burgers vector is colored red). The right insets show the indent sliced as illustrated in **a**.

### 3.5. Identification of defects at high temperatures

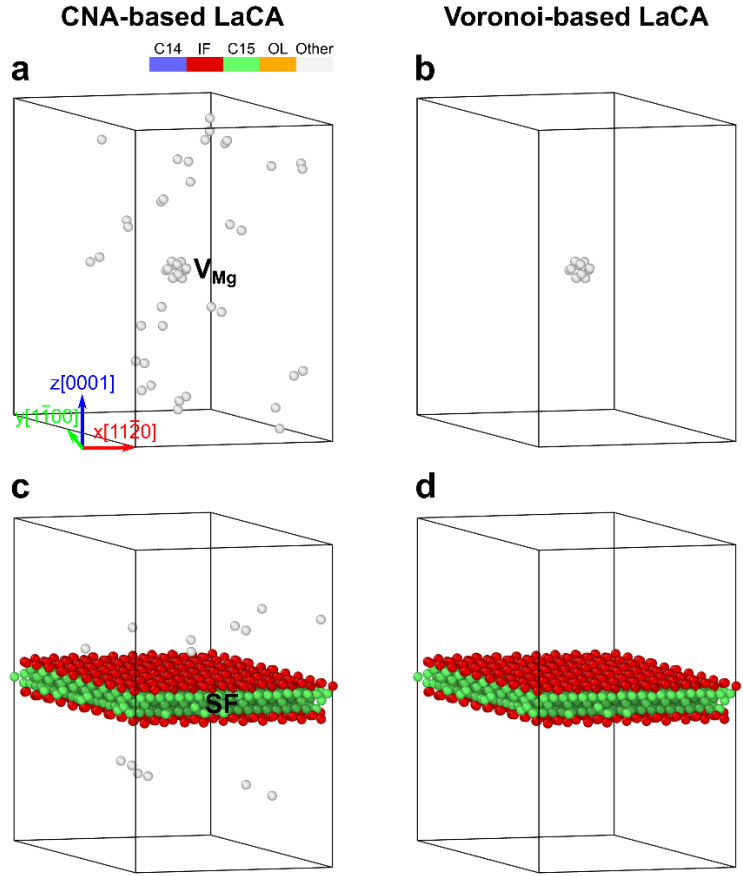

Figure 7: Defect structures of C14 Mg$_2$Ca atomistic configurations with **a,b** a Mg vacancy and **c,d** a stacking fault at 500 K. Characteristic atomic clusters were identified using **a,c** the modified CNA and **b,d** the Voronoi index. Colored by LaCA.

The sensitivity of the new method LaCA to thermal fluctuations was tested by analyzing the C14 Mg$_2$Ca atomistic configurations with defects at 500 K, which is approximately 50% of the melting temperature and the highest temperature considered experimentally for this alloy [19,45]. Details of the simulation method can be found in the Methodology section. The recognition of characteristic atomic clusters as the first step of the algorithm of LaCA can be carried out by the modified CNA or the Voronoi index. Figure 7a,c shows the defect atoms recognized by the CNA-based LaCA in the simulated samples at 500 K. Besides the atoms in the shell of the vacancy (see Figure 7a) and the stacking fault (see Figure 7c), a few atoms were also identified as Other due to the thermal noise. In contrast, the Voronoi-based LaCA is less sensitive to the thermal perturbations of the atomic coordinates. Only atoms in the nearest neighbors of the vacancy (Figure 7b) and the stacking fault (Figure 7d) were identified as defects.

Although the Voronoi index has certain limitations for the structural analysis of highly symmetric crystalline packings such as FCC and HCP [5,6], it shows better recognition rates of the Z12 and Z16 FK clusters than the modified CNA at elevated temperatures. The Voronoi index shows similar outcomes as the modified CNA at low temperatures (see Table S1), but it has a much higher computational cost (one order of magnitude slower) than the CNA. This makes the Voronoi-based LaCA unfavorable for large-scale configurations with millions of atoms. Therefore, we recommend to only employ the Voronoi index to analyze Laves phase structures at high temperatures and when it is not possible nor desired to reduce the thermal noise by other means such as time averaging.

## Conclusions

In this work, we present a new analysis method, Laves phase Crystal Analysis (LaCA), which can be used in atomistic simulations to identify the polytypes of C14 and C15 Laves phases and reveal common crystallographic and chemical defects of those. The atomistic structure is classified according to the neighbor list obtained from the close combination of a modified CNA and the CSP. By combining the modified CNA and DXA, the dislocation information can be extracted from plastically deformed Laves phase crystals. The new LaCA method now allows the automatic identification of

- hexagonal C14 and cubic C15 Laves phase crystal structures,
- point, linear and planar defects in the C14 and C15 Laves phases, including
    - basal slip via partial synchroshear dislocations
    - non-basal slip via perfect dislocations
    - basal stacking fault formed during synchroshear,
- mixtures of these defects from nanomechanical deformation tests of Laves phases.

Defects in the C36 Laves phase are currently not automatically identified, but detectable by eye through deviations from the periodic arrangement of the C14 and C15 structures that make up this phase. The implementation of the method, with a variable cutoff for the modified CNA and the permissible deviation in neighbor listing, makes LaCA robust against small perturbations of atomic coordinates. For atomistic structures with large atomic perturbations, i.e., at elevated temperatures, the Voronoi-based LaCA achieves a higher recognition rate of the local atomic environment than the CNA-based LaCA. Unlike the common structural identification methods, LaCA relies on the unique combination of FK clusters classification, centrosymmetry estimation and chemical order detection. By this, LaCA can not only identify complex crystal structures and structural defects, but also compositional defects like anti-site or anti-phase defects that are not captured by methods like CNA or CSP. Our novel method therefore paves the way for future atomistic studies aimed at understanding the deformation mechanisms of Laves and other TCP phases.

## Methodology

### Implementation

The new approach in the analysis of crystal structure and lattice defects in Laves phases, as one of the most abundant intermetallic structures, is currently available as a python program invoking a modified version of the open-source code OVITO.

*Variable cutoff of CNA*

The modified CNA is implemented in OVITO with adaptive cutoff, in which the optimal cutoff radius is determined automatically [5]. The adaptive cutoff $r_{adapt}$ is set to be halfway between the first and second neighbor shells:

$$r_{adapt} = \frac{1}{2}(r^{1st} + r^{2nd}), \tag{3}$$

and the ratio between $r_{adapt}$ and $r^{1st}$ is:

$$R = 0.5 + 0.5 \frac{r^{2nd}}{r^{1st}}. \tag{4}$$

The adaptive cutoff has to be modified to identify Z12 and Z16 FK clusters in Laves phase crystal structures. Figure 8 shows the radial distribution function of the prototype Laves phases and the C14 CaMg$_2$ Laves phase introduced in Section 3. For a B-centered Z12 FK cluster, the first neighbor shell consists of the first nearest B and A-type atoms within the cutoffs of $r_{B-B}^{1st}$ and $r_{A-B}^{1st}$, respectively (see Figure 8a). The second neighbor shell of the B-centered Z12 FK cluster consists of the second nearest B and A-type atoms within the cutoffs of $r_{B-B}^{2nd}$ and $r_{A-B}^{2nd}$, respectively. In our implementation, the cutoffs of the first and second neighbor shells of the Z12 FK cluster take the average values of $r_{B-B}^{1st/2nd}$ and $r_{A-B}^{1st/2nd}$. Therefore, the ratio between $r_{adapt}$ and $r^{1st}$ of the Z12 FK cluster is:

$$R_{Z12} = 0.5 + 0.5 \frac{r_{B-B}^{2nd} + r_{A-B}^{2nd}}{r_{B-B}^{1st} + r_{A-B}^{1st}}. \tag{5}$$

Similarly, for an A-centered Z16 FK cluster:

$$R_{Z16} = 0.5 + 0.5 \frac{r_{A-A}^{2nd} + r_{A-B}^{2nd}}{r_{A-A}^{1st} + r_{A-B}^{1st}}. \tag{6}$$

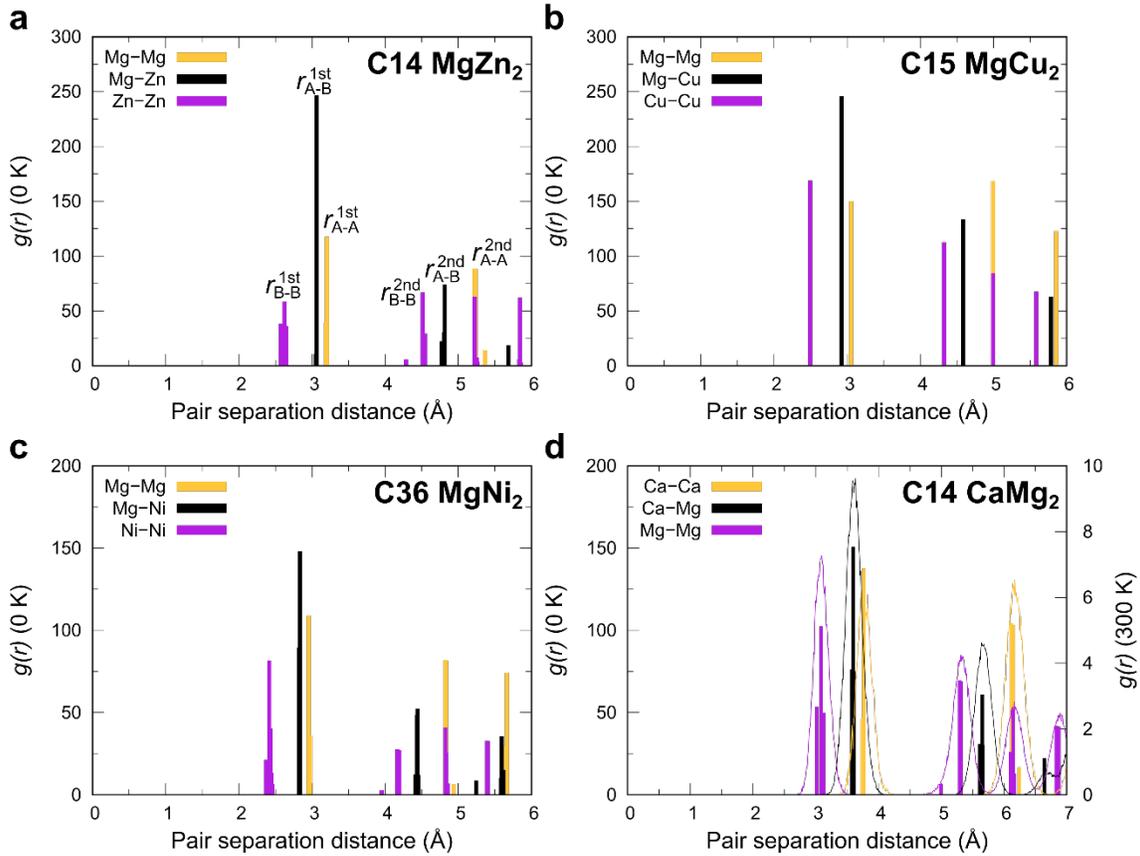

Figure 8: Radial distribution function of the prototype **a** C14 MgZn$_2$, **b** C15 MgCu$_2$ and **c** C36 MgNi$_2$ Laves phase atomistic structures. **d** Radial distribution function of the C14 CaMg$_2$ atomistic sample. The curves indicate the distribution function at 300 K.

Taking the radial distribution function of the C14 CaMg$_2$ atomic sample ($a$=6.262 Å and $c$=9.991 Å) modelled using the interatomic potential [27] and the prototype Laves phases C14 MgZn$_2$ ($a$=5.221 Å and $c$=8.567 Å) [46], C15 MgZn$_2$ ($a$=7.060 Å) [47] and C36 MgNi$_2$ ($a$=4.824 Å and $c$=15.826 Å) [46], $R_{Z12}$=1.32 and $R_{Z16}$=1.30. Based on the calculation, we set the default $R$ values in our implementation as 1.32 for Z12 FK cluster and 1.30 for Z16 FK cluster. It is worth noticing that the choice of cutoff is somewhat arbitrary in the original algorithm [5] and a more recent work proposed using two-third of the distance instead of halfway [48].

*CSP threshold*

To classify B1- and B2-type atoms, an appropriate threshold should allow for small perturbations due to thermal noises and elastic strains. According to the distribution of probability density of the CSP values of all central atoms of Z12 FK clusters in C14 Mg$_2$Ca Laves phase crystals as well as in the prototype Laves phases, we set the threshold at 5. With a CSP value greater than 5, the atom is identified as B2-type, otherwise, the atom is identified as B1-type.

**Simulation methods**

Atomistic simulations were carried out using the software package LAMMPS [49]. The code Atomsk [50] was used to construct the atomistic configurations of the Laves phases. The interatomic interactions were modelled using the modified embedded atom method (MEAM) potential from Kim *et al*. [27] published in 2015. Energy minimization was performed using a combination of the conjugate gradient (CG) algorithm with box relaxation and the FIRE algorithm [51,52] until the force norm was below $10^{-8}$ eV/Å.

*Typical crystallographic defects*

A C14 (C15) $Mg_2Ca$ sample with dimensions of 6.3 × 5.4 × 10.0 $nm^3$ (6.4 × 4.9 × 12.0 $nm^3$) containing 12,000 (13,824) atoms was constructed. Energy minimization was performed after sample generation with box relaxation in all directions.

The point defects were introduced by removing an atom (vacancy) or swapping atom types (anti-site) in the C14 sample with periodic boundary conditions (PBCs) in all directions. Energy minimization was performed on the atomic samples with box relaxation in all directions.

The stacking faults were introduced by shifting one half of the crystal along one partial Burgers vector across the slip planes. The twin boundary in C15 was constructed by merging two mirror symmetric crystals along the [111] plane. PBCs were removed in the direction perpendicular to planar faults. The samples were allowed to relax in the directions perpendicular to the planar faults.

The free surfaces were created by removing PBCs in the surface normal direction. Energy minimization was performed on the atomistic samples with box relaxation in the directions orthogonal to the surface normal direction.

The synchro-Shockley dislocations were obtained from our previous study on the synchroshear mechanism using nudged-elastic band (NEB), where details of the simulations can be found [28].

*Nanoindentation*

The dimensions of the atomic sample are 61.5 × 31.8 × 59.9 $nm^3$ with around 4.3 Mio atoms. PBCs were applied along x-[$2\bar{1}\bar{1}0$] and z-[0001] directions. A layer of atoms with thickness of 1 nm at the bottom of the sample was rigidly fixed. A spherical indenter exerted a force of magnitude $F(r) = -K(r-R)^2$ on each atom, where $K$ is the specified force constant ($K$ = 10 eV·Å$^{-3}$), $r$ is the distance from the center of the indenter to the atom, and $R$ is the radius of the indenter ($R$ = 10 nm). The indenter was displaced along the y-[$\bar{1}2\bar{1}0$] direction toward the sample with a velocity of 10 m/s and a maximum indent depth of 5 nm. The microcanonical (NVE) ensemble was applied during the nanoindentation tests with a time step of 1 fs. Energy minimization was performed on the indented configuration, while keeping the indenter at the maximum depth.

*At elevated temperatures*

The procedures of sample generation and energy minimization were conducted in a similar way as described above. The atomic samples were heated up to 500 K with a heating rate of $10^{12}$ K/s and stress relaxation in periodic directions within the isothermal-isobaric (NPT) ensemble. Then the samples were annealed at 500 K for 50 ps. Nosé-Hoover thermostat and barostat were used to control the temperature and pressure [53]. A similar approach was applied to heat up the defect-free bulk C14 $Mg_2Ca$ sample to 300 K.

## Acknowledgements

The authors acknowledge financial support by the Deutsche Forschungsgemeinschaft (DFG) through the projects A02, A05 and C02 of the SFB1394 Structural and Chemical Atomic Complexity – From Defect Phase Diagrams to Material Properties, project ID 409476157. This project has received funding from the European Research Council (ERC) under the European Union's Horizon 2020 research and innovation programme (grant agreement No. 852096 FunBlocks). EB gratefully acknowledges support from the German Research Foundation (DFG) through projects C3 of the collaborative research centre SFB/TR 103. Simulations were performed with computing resources granted by RWTH Aachen University under project (rwth0591), by the EXPLOR center of the Université de Lorraine and by the GENCI-TGCC (Grant 2020-A0080911390). Z.X. wants to thank Dr.-Ing. Wei Luo (RWTH Aachen) for fruitful discussions.

## Data availability

We have implemented our new method LaCA for use by other researchers and made our code and relevant atomic configurations available online [54].

## Compliance with ethical standards

Conflict of interest The authors declare no confict of interests.

# Supplementary material for:

# Laves Phase Crystal Analysis (LaCA): Atomistic Identification of Lattice Defects in C14 and C15 Topologically Close-Packed Phases

Zhuocheng Xie[a], Dimitri Chauraud[b], Erik Bitzek[b], Sandra Korte-Kerzel[a],

Julien Guénolé[a,c,d]

[a] Institute of Physical Metallurgy and Materials Physics, RWTH Aachen University, 52056 Aachen, Germany
[b] Department of Materials Science & Engineering, Institute I: General Materials Properties, Friedrich-Alexander-Universität Erlangen-Nürnberg, 91058 Erlangen, Germany
[c] Université de Lorraine, CNRS, Arts et Métiers ParisTech, LEM3, 57070 Metz, France
[d] Labex Damas, Université de Lorraine, 57070 Metz, France

Table S1: Comparison of the modified CNA and the Voronoi index in identification of characteristic atomic clusters in simulated C14 $Mg_2Ca$ atomistic structures at different temperatures.

| Simulation | Temperature | Recognition rate (modified CNA) | | | Recognition rate (Voronoi index) | | |
|---|---|---|---|---|---|---|---|
| | | Z12 | Z16 | Other | Z12 | Z16 | Other |
| NEB in *Section 3.2* | 0 K | 54.9% | 26.6% | 18.5% | 55.5% | 26.7% | 17.8% |
| Pillar compression in *Section 3.3* | 50 K | 26.6% | 12.5% | 60.9% | 26.4% | 12.6% | 61.0% |
| Nanoindentation in *Section 3.4* | 0 K | 65.1% | 32.2% | 2.7% | 65.4% | 32.2% | 2.4% |
| Mono-vacancy in *Section 3.5* | 500 K | 66.6% | 33.0% | 0.4% | 66.6% | 33.3% | 0.1% |

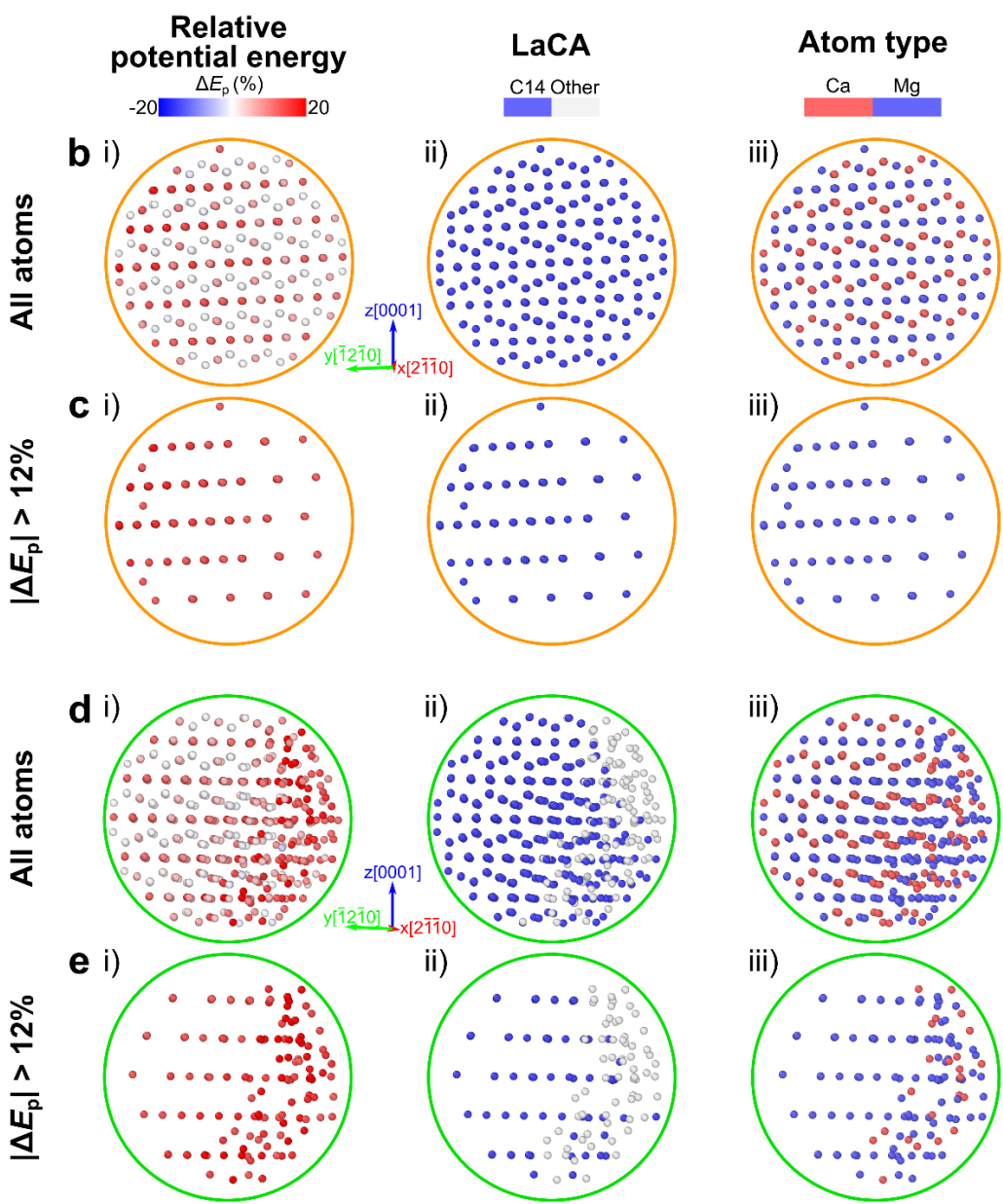

Figure S1: **a** Deformation zone beneath the indent in the nanoindentation test on the C14 Mg$_2$Ca atomistic structure. Left: Energy filtered configurations where only atoms with $\Delta E_p$ > 12% are visible. Right: Only atoms with $|\Delta E_p|$ > 12% & identified as C14 structure by LaCA are visible. Atoms are colored by $\Delta E_p$. Surface mesh is colored grey. **b, d** Atoms in the regions of interest marked in **a** where Mg atoms show high $\Delta E_p$ and local structure identified as C14. **c, e** Only atoms with $|\Delta E_p|$ > 12% in the regions of interest marked in **a** are visible. i) Colored by $\Delta E_p$. ii) Colored by LaCA. iii) Colored by atom type.

**Comment**: This figure shows that all high-energy atoms due to plastic deformation have been identified as defects by LaCA. The atoms not identified as defect by LaCA (i.e., identified as C14) but yet showing high $\Delta E_p$ are close to defects. In particular, Mg atoms that are highly elastically strained were correctly identified as C14 structure by LaCA, even though there high $\Delta E_p$ would give the false impression that they are forming defects.